**Radiation-hardened and Repairable MoS₂ Field Effect Devices with Polymer Solid Electrolyte Gates**

*Di Chen, Jiankun Li, Zheng Wei, Xinjian Wei, Maguang Zhu, Jing Liu, Guangyu Zhang, Zhiyong Zhang, and Jian-Hao Chen\**

Dr. D. Chen, Dr. X.-J. Wei, Prof. J.-H. Chen
International Center of Quantum Material, School of Physics, Peking University, Beijing 100871, China
E-mail: chenjianhao@pku.edu.cn

Dr. D. Chen, Dr. J. K. Li, Dr. X.-J. Wei, Dr. J. Liu, Prof. J.-H. Chen
Beijing Academy of Quantum Information Sciences, Beijing 100193, China

Z. Wei, Prof. G. Y. Zhang
Institute of Physics, Chinese Academy of Sciences, Beijing 100190, China

Dr. M. Zhu, Prof. Z. Zhang, Prof. J.-H. Chen
Key Laboratory for the Physics and Chemistry of Nanodevices, Peking University, Beijing 100871, China

Prof. J.-H. Chen
Interdisciplinary Institute of Light-Element Quantum Materials and Research Center for Light-Element Advanced Materials, Peking University, Beijing

As human activities expand into naturally or man-made radiation-prone environment, the need for radiation-hardened (Rad-Hard) electronic hardware surged. The state-of-the-art silicon-based and two-dimensional (2D) materials based Rad-Hard transistors can withstand up to 1 Mrad (Si) of total ionization dose (TID), while higher TID tolerance is being heatedly sought after. Here we present few-layer MoS₂ Rad-Hard field-effect transistors (FETs) with polymer solid electrolyte (PSE) gate dielectrics. The MoS₂ PSE-FETs exhibit a TID tolerance of up to 3.75 Mrad (Si) at a dose rate of 523 rad (Si) s⁻¹ and can be repaired with a moderate thermal annealing at 100 °C for 5 minutes. Combining the excellent intrinsic radiation tolerance and the reparability, the MoS₂ PSE-FETs reach a TID tolerance of up to 10 Mrad (Si). Complementary metal–oxide–semiconductor (CMOS)-like MoS₂ PSE-inverters have been built and show similar high radiation tolerance. Furthermore, the feasibility of wafer-scale Rad-Hard PSE-inverter array has been demonstrated using chemical vapor deposition (CVD) grown monolayer $MoS_2$. Our studies uncover the potential of 2D materials based PSE devices in future Rad-Hard integrated circuits (ICs).

## 1. Introduction

With the continuous progress of human's aerospace technology, nuclear physics, accelerator science and radiation therapy technology, the demand for Rad-Hard ICs has been booming. In the last 50 years, benefiting from the general reduction in the thickness of dielectric layers driven by Moore's law and the development of radiation-tolerant technologies, Si-based metal–oxide–semiconductor (MOS) devices and ICs have already been able to survive in typical space applications.[1] However, operation of modern electronics in stellar exploration missions, nuclear reactors, particle accelerators and other harsh radiation environments calls for higher radiation tolerance. For Si-based ICs, special designs with redundancy are proposed to improve the circuits' overall radiation tolerance[2]; in the single device level, with the development of new semiconductor materials, attempts to increase the radiation tolerance of the channel in FETs have been reported using oxide semiconductors, carbon nanotubes and 2D materials.[3]

In particular, 2D materials have been proposed as compelling candidates for the post-silicon age.[4] Hence, it is of great significance to study radiation-hardening technology of 2D materials based devices. In the family of 2D materials, $MoS_2$ is a promising semiconductive channel material, which has shown remarkable characteristics in realizing flexible transistors, optoelectronic devices and logic gates.[5, 6] Previous studies have reported proton, X-ray, electron and ion-beam irradiation of $MoS_2$ FETs with conventional gate dielectrics including $SiO_2$, $ZrO_2$ and $HfO_2$, which displays an obvious threshold voltage shift and rapidly growing leakage current.[7, 8] Other studies indicate that the $MoS_2$ material is extraordinarily radiation resilient, while the oxide dielectric damage accounts for the majority of radiation induced failure of the FETs.[9] Ionic liquids or gels have been used as high-efficiency gates for $MoS_2$ FETs,[6, 10, 11] while the radiation resistance of these devices has not been studied. In particular, the radiation tolerance of the technologically more attractive solid ionic electrolyte gated 2D materials devices is still unexplored.

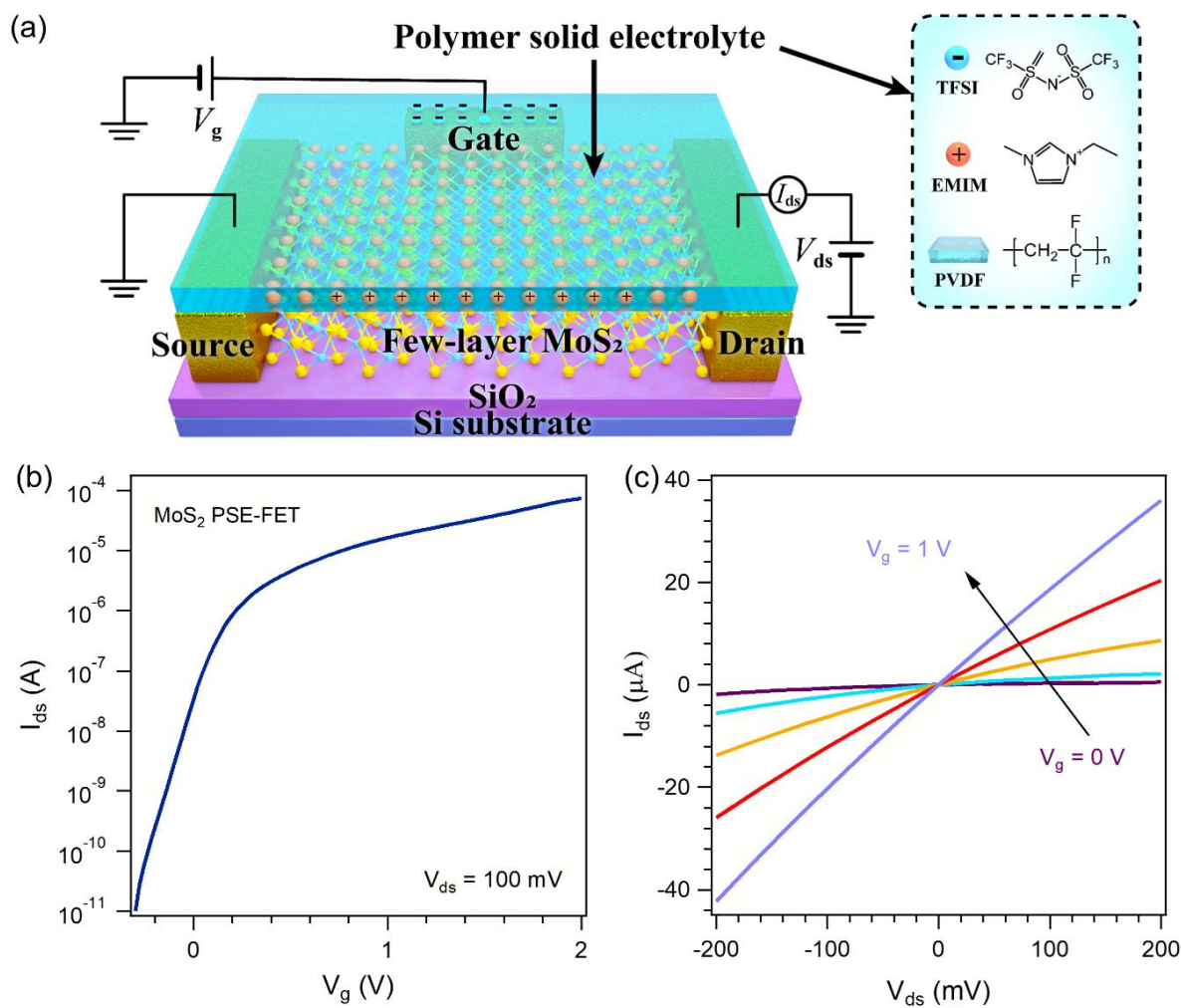

**Figure 1.** Characterization of a typical $MoS_2$ PSE-FET. a) Schematic view of a $MoS_2$ PSE-FET built on a $SiO_2$/Si substrate; the composition of the PSE is shown in the right section. b) Room temperature transfer characteristics for the $MoS_2$ PSE-FET with 100 mV bias voltage. c) $I_{ds}$-$V_{ds}$ curves acquired for $V_g$ values of 0, 0.25, 0.5, 0.75 and 1 V.

In this article, we report the realization of PSE gated few-layer $MoS_2$ FETs and inverters with high radiation tolerance and reparability after radiation damage. Few-layer $MoS_2$ exhibits large on-off ratio, atomic channel thickness and good electrical contact to metal electrodes,[12] making them suitable for integrated circuits operating in radiation-prone environments. PSE gates provide high gate efficiency by allowing the formation of electric double layer (EDL) at the surface of the $MoS_2$ channel with the stability of the solid polymer. In this study, we found that $MoS_2$ PSE devices have excellent radiation tolerance and fast self-recovery capability with an annealing process at a moderate temperature of 100 °C for 5 min. The $MoS_2$ PSE-FETs and inverters made of mechanically exfoliated $MoS_2$ exhibit a radiation tolerance up to 3.75 Mrad (Si) and 10 Mrad (Si) at a dose rate of 523 rad (Si) $s^{-1}$, without and with the annealing procedure, respectively. The feasibility of wafer-scale Rad-Hard PSE-inverter array has also been demonstrated using CVD grown monolayer $MoS_2$, as the inverters function well except for an average positive shift of 0.386 V for the transition voltage $V_{tr}$ after 0.5 Mrad (Si) of irradiation at a dose rate of 523 rad (Si) $s^{-1}$.

## 2. Design of radiation-hardened FETs with PSE gates

The design of Rad-Hard FETs is illustrated schematically in **Figure 1a**. The few-layer $MoS_2$ channel is obtained by mechanical exfoliation, and 5nm chromium /50 nm gold electrodes are defined with standard electron beam lithography. The PSE layer covering the channel material is composed of polyvinylidene fluoride (PVDF) and the ionic liquid 1-ethyl-3-methylimidazolium bis[(trifluoromethyl)sulfonyl] imide (EMIMTFSI). Details of the processes for preparing the PSE layer and fabricating the FETs are described in the Experimental Section and in Supplementary Figure S1. Our polymer electrolyte films behave like solids mechanically, while their ionic conductivity closely resembles that of the ionic liquid.[13] Under a gating electric field, the mobile ions redistribute inside the solid electrolyte, which form an EDL at the electrolyte/semiconductor interface, resulting in a high gate capacitance.

**Figure 1b** shows the transfer characteristics, e.g., drain-source current $I_{ds}$ vs. gate voltage $V_g$, for a typical $MoS_2$ PSE-FET with 100 mV bias voltage $V_{ds}$. The device works as a typical n-FET with an on-current $I_{on}$ of 73.6 μA μm$^{-1}$ and current on/off ratio $I_{on}/I_{off} > 10^6$ for $V_g$ from -1V to 2V. Here we assign the off-state current $I_{off} = 10$ pA, as lower $I_{ds}$ could be obscured by the minuscule gate leakage. From the $I_{ds}$ vs. $V_g$ curve we deduce a subthreshold slope $SS \sim$ 89.6 mV dec$^{-1}$ and transconductance $g_m \sim 82.4$ μS μm$^{-1}$ at $V_{ds} = 100$ mV. **Figure 1c** shows $I_{ds}$-$V_{ds}$ curves recorded for different values of $V_g$. The linear dependence of $I_{ds}$ on $V_{ds}$ indicates that the electrical contacts are Ohmic. Thus, we have realized an FET with few-layer $MoS_2$ as a conductive channel and PSE as gate dielectric that show similar high gate efficiency as ionic liquid gates.[11, 14] Furthermore, our devices are free from the disadvantage of ionic liquid electrolyte, such as the mechanical instability of the liquid gate dielectric and the sensitivity to the presence of moisture.[15] Our PSE devices are stable in air for at least 7 days, as shown in the stability of both the threshold voltage $V_{th}$ and $I_{on}/I_{off}$ before and after storage in air (see Supplementary Figure S2).

## 3. TID effects of PSE gated devices with 2D materials channel

In this section, the TID effects of PSE gated devices with $MoS_2$ channel are studied; as a comparison, TID effects of graphene PSE-FETs are also investigated. Ionizing radiation includes α-particle, β-particle, energetic protons, γ-rays and X-rays.[16] High energy photons such as γ-rays and X-rays could cause the most ionization damage in electronic circuits with modest protection, since massive particles such as α-particle, β-particle and medium-energy protons (e.g. with kinetic energy ~1keV as in solar wind) can be shielded relatively easily.[17] We shall concentrate our study on a radiation environment realized by a $^{60}$Co γ-ray source. Here, the absorbed dose and dose rate are defined as the amount of energy deposited by ionizing radiation in silicon (Si). In the following text, the "(Si)" will be omitted for simplicity.

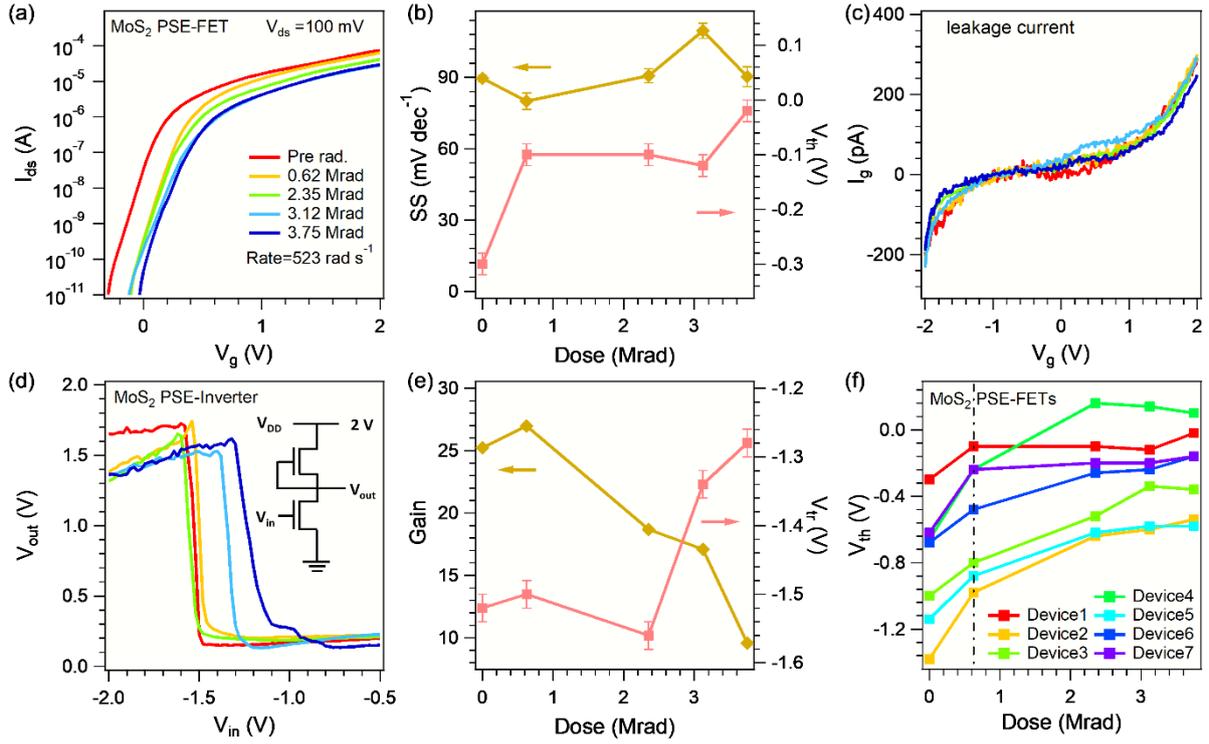

**Figure 2.** TID-dependent properties of $MoS_2$ PSE-FETs and inverters. a) Transfer characteristics of the $MoS_2$ PSE-FET at $V_{ds} = 100$ mV with different TIDs. b) Subthreshold slope $SS$ and threshold voltage $V_{th}$ for the $MoS_2$ PSE-FET as a function of TID. c) Leakage current of the $MoS_2$ PSE-FET as a function of $V_g$ with different TIDs. d) VTC curves of the $MoS_2$ inverter for $V_{dd} = 2$ V with different TIDs. Inset: schematic drawing of the inverter's electronic circuit. e) Peak gains and $V_{tr}$ of the $MoS_2$ inverter at $V_{dd} = 2$ V as a function of TID. f) Threshold voltages $V_{th}$ of 7 $MoS_2$ PSE-FETs as a function of TID. The dashed line indicates TID = 0.62 Mrad. Error bars for $SS$ and Gain represent one standard deviation of the fitting, while error bars for $V_{th}$ and $V_{tr}$ represent the voltage scanning step size.

**Figure 2a** shows the transfer curves of a typical $MoS_2$ PSE-FET in the pristine state and after five different TIDs. For radiation doses up to 3.75 Mrad, only slight changes occur in its transfer curves. It is worth noting that the large dose rate of 523 rad s$^{-1}$ we used here simulates a much tougher radiation environment than that in outer-space, where the dose rate is of order 0.01 rad s$^{-1}$.[16, 18] In **Figure 2b**, $SS$ changes slightly from 89.6 mV dec$^{-1}$ to 90.3 mV dec$^{-1}$ and $V_{th}$ shifts from -0.3 V to -0.02 V, and as shown in **Figure 2c**, the small gate leakage currents show almost no change with different TIDs. Although the slight positive $V_{th}$ shift indicates radiation induced electron-trapping near the PSE-semiconductor interface, the $MoS_2$ PSE-FET presents excellent radiation tolerance.

To characterize the radiation tolerance of ICs based on MoS$_2$ PSE-FETs, a PSE-inverter is constructed using two MoS$_2$ PSE-FETs, with its circuit diagram shown in the inset of **Figure 2d**. Voltage transfer characteristics (VTC) curves of the PSE-inverter under different TIDs are shown in **Figure 2d**. Obvious performance change was not observed until after 3 Mrad of TID. Both $V_{tr}$ and gain dependence on TID are extracted and summarized in **Figure 2e**. With irradiation of up to 3.75 Mrad, $V_{tr}$ varies only from -1.52 V to -1.28 V. The inverter's voltage gain, on the other hand, has a more appreciable decrease of 70% of its original value at 3.75 Mrad TID. Based on the positive shift of $V_{th}$ for the MoS$_2$ PSE-FETs and similar positive shift of $V_{tr}$ for the MoS$_2$ PSE-inverter due to irradiation, we can draw a natural conclusion that modification of $V_{th}$ of the two PSE-FETs constituting the inverter results in the modification of $V_{tr}$ of the inverter. The dependence of voltage gain of the inverter on TID is jointly caused by $SS$ degradation induced by interfacial traps and the difference in the change of $V_{th}$ of the two PSE-FETs.

**Figure 2f** shows $V_{th}$ vs. TID for several MoS$_2$ PSE-FETs. An interesting observation was that $V_{th}$ of MoS$_2$ PSE-FETs exhibits larger positive shifts upon the first 0.62 Mrad of irradiation and slower shifts for subsequent doses. This suggests a general way to increase TID tolerance could simply be a radiation pretreatment of the devices *before* putting them to service. To test whether this pretreatment is generally applicable or not, we also made graphene PSE-FETs and tested their TID response as shown in Supplementary Figure S3. Indeed, similar trends were observed for the bipolar graphene PSE-FETs; the transfer curves and the gate voltage of minimum conductivity $V_{min}$ (correspondent to $V_{th}$ in MoS$_2$ PSE-FETs) only change slightly for TID exceeding 0.62 Mrad. Such behavior in MoS$_2$ PSE-FETs and graphene PSE-FETs indicates that the saturating response of $V_{th}$ or $V_{min}$ on TID is likely caused by the PSE dielectric, meaning that it could be generally applied to other 2D materials PSE-FETs.

Such saturation behavior could be understood as following. Through a process similar to Compton scattering, the γ-ray irradiation generates electron-ion pairs in the PSE.[19] Some ions combine with electrons again, while others form charge traps along their diffusive paths. With the accumulation of trapped charges induced by irradiation, the activation energy of electron-ion pairs generation will be elevated,[19] resulting in a reduction in efficiency for subsequent γ-ray to cause damage in PSE; meanwhile, radiation-induced neutralization effect might also be enhanced with higher ion density created by preceding irradiation.[20] Both effects lead to a slower $V_{th}$ or $V_{min}$ shift for subsequent radiation.

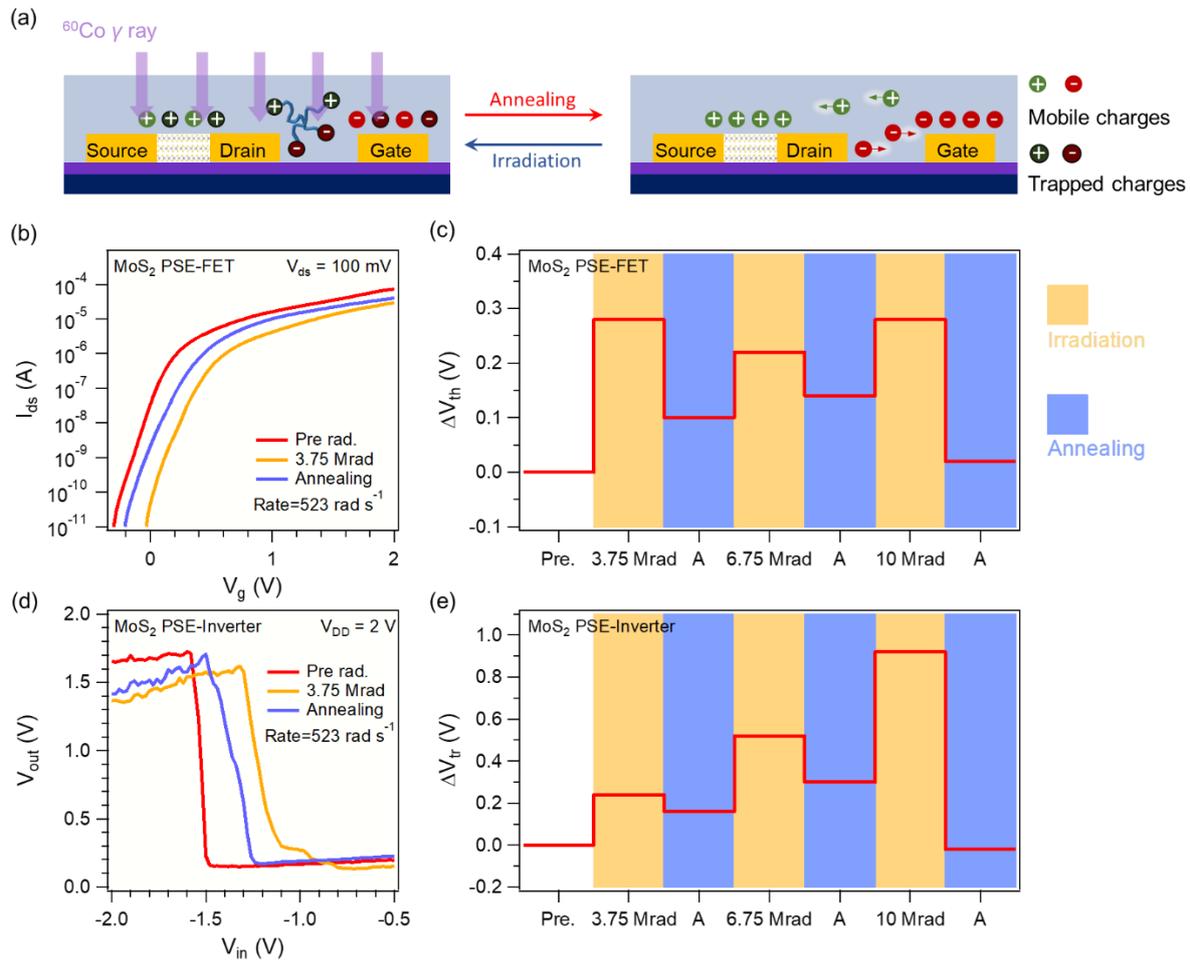

**Figure 3.** Radiation damage recovery of MoS$_2$ PSE-FET and inverter via annealing. a) Schematics illustrating radiation recovery processes of the polymer electrolyte film via annealing. b) Annealing-induced transfer characteristics recovery of the MoS$_2$ PSE-FET. c) $V_{th}$ shifts during multiple cycles of irradiation and annealing of the MoS$_2$ PSE-FET. d) Annealing-induced VTC property recovery of the MoS$_2$ PSE-inverter. e) $V_{tr}$ shifts during multiple cycles of irradiation and annealing of the MoS$_2$ PSE-inverter. The tick labels mark the total irradiation dose and "A" represents after an annealing process of 100 °C for 5 min.

**4. Damage recovery via annealing**

In previous sections, we have demonstrated that $MoS_2$ PSE-FETs and inverters have a radiation tolerance over 3 Mrad at a dose rate of 523 rad s$^{-1}$, which is considerably higher than the radiation tolerance of around 1 Mrad for previously reported Si-based ICs.[21] Nevertheless, further irradiation will still lead to failure of $MoS_2$ PSE devices. Such "failure" is defined according to standards of commercial radiation-hardened Si-based ICs, as the change of threshold voltage $\Delta V_{th}$ or $\Delta V_{tr} > 0.3V$.[22] For Si-based ICs, thermal annealing processes of about 300 to 900 °C for 4 to 30 minutes,[23, 24] or around 100 °C for several hours have been employed to partially recover the initial threshold voltage.[25]

In PSE devices, we found that a much milder condition could induce a large recovery effect. Such behavior likely arises from the large mobility of the conduction ions in PSE.[26] The recovery process can be illustrated in **Figure 3a**, as the trapped charges are neutralized by thermally excited electrons by heating. In addition, higher temperature accelerates the dynamic recovery of the polymer matrix and the neutralization of radicals.[27] In this way, the device performance could be partially restored, increasing the overall TID tolerance of such devices. **Figure 3b** shows the transfer characteristics of a typical $MoS_2$ PSE-FET in the pristine, irradiated and annealed state. The $MoS_2$ PSE-FET exhibits $\Delta V_{th}$ = 0.28 V after 3.75 Mrad TID irradiation and $\Delta V_{th}$ = -0.18 V after annealing at 100 °C for 5 min, amounting to a 64% recovery. This annealing temperature is quite low compared with that of the Si-based ICs, which is over 300 °C with comparable time.[23] The annealing induced recovery of a typical $MoS_2$ PSE-inverter is similar to that of a single $MoS_2$ PSE-FET, in which $V_{tr}$ exhibits a positive shift after irradiation and a partial recovery after annealing.

Multiple radiation-recovery cycles were performed and the effective $\Delta V_{th}$ ($\Delta V_{tr}$) relative to the original threshold (transition) voltage is displayed in **Figure 3c** (**Figure 3e**) for the MoS$_2$ PSE-FET (inverter), respectively. Both the MoS$_2$ PSE-FET and inverter are able to recover through annealing after every radiation cycle. For the MoS$_2$ FET, $\Delta V_{th}$ are always within 0.3 V during the three cycles, which indicates a combined radiation tolerance of up to 10 Mrad TID if the annealing procedure is included. For the MoS$_2$ inverter, $\Delta V_{tr}$ exceeds 0.3 V at 6.75 Mrad TID, which is due to the slight difference between the $V_{th}$ vs. TID curves and their recovery conditions of the pull-up and pull-down MoS$_2$ PSE-FETs constituting the inverter. More data on the radiation-recovery cycles of the MoS$_2$ and graphene PSE-FETs can be found in Supplementary Figure S4 and Figure S5, respectively. From the above discussion, we conclude that annealing guarantees at least a partial recovery of the devices and could be used to greatly extend the lifetime of MoS$_2$ PSE devices in radiation-prone environment.

## 5. TID effect of wafer-scale MoS$_2$ PSE-inverter array

Towards practical application of radiation-hardened scaled-up electronics, the TID effect of wafer-scale MoS$_2$ PSE-inverter array was studied. The wafer-scale monolayer MoS$_2$ crystal is grown on sapphire by a three-temperature-zone CVD system.[28] **Figure 4a** shows an optical micrograph of the batch-fabricated MoS$_2$ PSE-FETs array and the zoomed out micrograph is shown in the inset of **Figure 4a**. For simplicity, one CVD MoS$_2$ PSE-FET in series with a 10 M$\Omega$ resistor are used to form a CVD MoS$_2$ PSE-inverter as depicted in the inset of **Figure 4b**. We mapped the VTC curves of a 5×10 array of MoS$_2$ PSE-inverters (**Figure 4b**), with their initial $V_{tr}$ shown in **Figure 4c**. Before radiation, $V_{tr}$ in the inverter array has an initial device-to-device variation from 0 V to -1.22 V. Then, we exposed the inverter array to 0.5 Mrad of irradiation at a rate of 523 rad s$^{-1}$ and the resulting histogram of $\Delta V_{tr}$ is shown in **Figure 4d**. Except for an average $\Delta V_{tr} \approx 0.386$ V, the PSE-inverters does not show much changes after irradiation: for input voltages corresponding to logic 0, the output voltage is still very close to the supply voltage of $V_{dd} = 2$ V (see Supplementary Figure S6 for the VTC curves). A comparison between exfoliated and CVD MoS$_2$ PSE-inverters reveals that: 1) both types of the devices have positive $\Delta V_{tr}$ after irradiation; 2) different from exfoliated MoS$_2$ PSE-inverters, 30% of the 50 CVD MoS$_2$ PSE inverters has $\Delta V_{tr} < 0.3$V under 0.5 Mrad TID at a rate of 523 rad s$^{-1}$. We also perform another radiation experiment on CVD MoS$_2$ PSE-inverter array with higher TID of 4.5 Mrad and at lower radiation rate of 60.7 rad s$^{-1}$. Similar to the results shown in **Figure 4**, except for an average $\Delta V_{tr} \approx 0.547$ V, no obvious change can be observed in the VTC curves of the inverters. The VTC curves and the histogram of $\Delta V_{tr}$ of the CVD MoS$_2$ PSE-inverter array for the 4.5 Mrad TID and 60.7 rad s$^{-1}$ radiation experiment are shown in Supplementary Figure S7a and S7b, respectively.

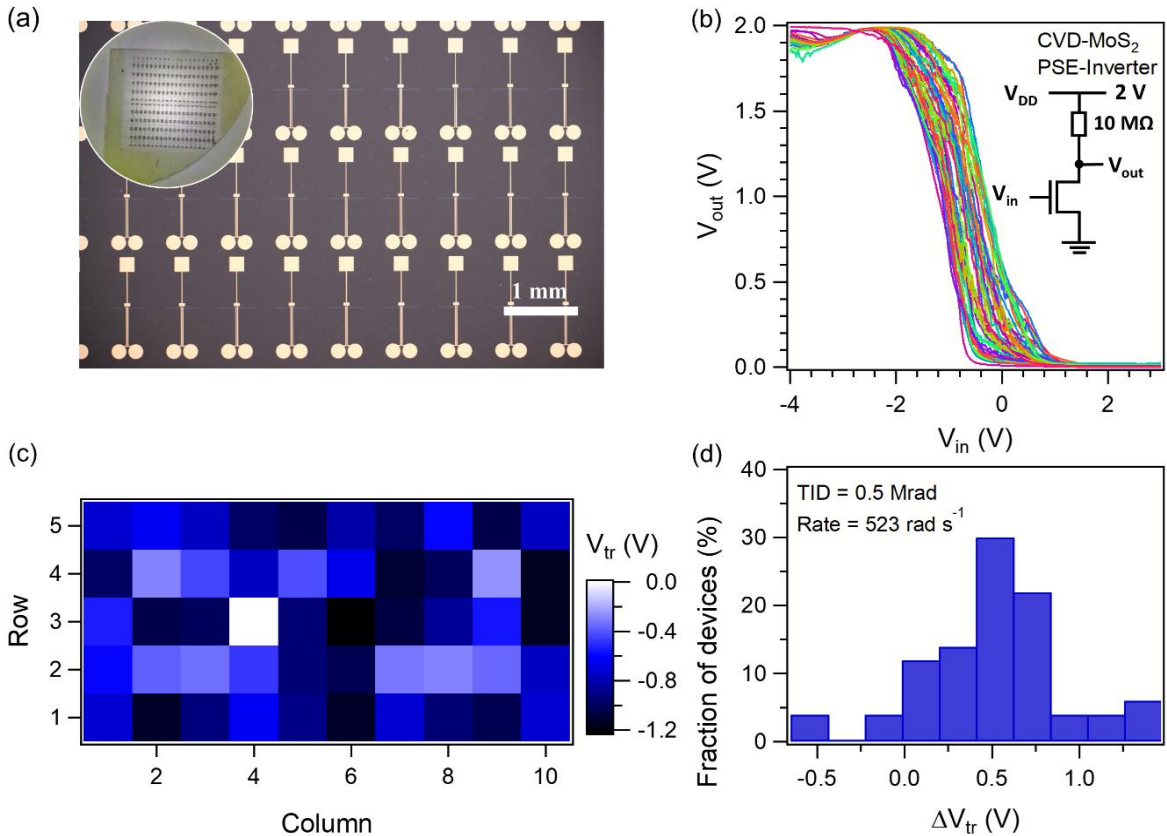

**Figure 4.** TID-dependent properties of wafer-scale CVD grown MoS$_2$ PSE-inverters. a) Optical images of the batch-fabricated MoS$_2$ PSE-FETs array on sapphire wafer. b) VTC curves of the 5×10 array of MoS$_2$ inverters before irradiation. Inset: schematic drawing of the inverter's electronic circuit. c) $V_{tr}$ mapping of the 5×10 array of MoS$_2$ PSE-inverters before irradiation. d) Histogram of statistical $\Delta V_{tr}$ data after 0.5 Mrad (rate = 523 rad s$^{-1}$) radiation for the 5×10 array of MoS$_2$ PSE-inverters.

For samples used in this study, the CVD grown $MoS_2$ PSE-FETs exhibit a lower *SS* and mobility than the mechanically exfoliated $MoS_2$ PSE-FETs (see Supplementary Figure S8). One reason is that the exfoliated $MoS_2$ is multilayer with smaller bandgap, leading to higher mobility compared with CVD grown monolayer $MoS_2$.[29] Another reason may lie in the additional domains and defects that might present in CVD grown monolayer $MoS_2$. Defects and sulfur vacancies can absorb oxygen during irradiation and act as surface electron traps, making the CVD grown $MoS_2$ more vulnerable to irradiation.[8, 30] It is reasonable to expect that with improved CVD sample quality control, wafer-scale $MoS_2$ ICs with large radiation tolerance could be realized in the near future.

## 6. Conclusion

Exceptional irradiation resistance is demonstrated in $MoS_2$ polymer solid electrolyte FETs and inverters exposed to $^{60}Co$ γ-rays. The mechanically exfoliated $MoS_2$ polymer solid electrolyte FETs and inverters work well under TID of up to 3.75 Mrad at a high dose rate of 523 rad s$^{-1}$, except for positive threshold or transition voltage shifts within 0.3 V. Moreover, the radiation damage is reparable and the devices can partially recover via moderate thermal annealing of 100 °C for 5 min. With annealing, our $MoS_2$ PSE-FETs reached a TID tolerance up to 10 Mrad. Wafer-scale radiation-hardened $MoS_2$ inverter array has also been demonstrated using CVD $MoS_2$. The inverters function well except for an average positive shift of 0.386 V of the transition voltage with 0.5 Mrad TID at a high dose rate of 523 rad s$^{-1}$. Our study is an important first step towards the application of large-scale radiation-hardened electronic circuits made from two-dimensional van der Waals semiconductors.

## 7. Experimental Section

*Materials*: The few-layer $MoS_2$ and graphene were mechanically exfoliated onto bulk Si substrates with 280 nm $SiO_2$. The wafer-scale monolayer $MoS_2$ on sapphire was grown by a three-temperature-zone CVD system. Sapphire wafers were annealed in an oxygen atmosphere at 1000 °C for 4 hours prior to the growth to form an atomically flat surface for subsequent $MoS_2$ growth. A typical growth lasts for ∼40 min under the pressure of ∼1 Torr.

*Devices Fabrication and Characterization*: For mechanically exfoliated $MoS_2$ and graphene devices, Au (50nm)/Cr (5nm) electrodes were patterned into a Hall bar configuration with an additional gating electrode by a standard e-beam lithography process. For CVD-grown $MoS_2$ devices, a laser direct writing system was used to fabricate the patterns on the wafer-scale monolayer $MoS_2$. Then through oxygen plasma etching, Au/Ti/Au (2/2/50 nm) electron beam evaporation and lift-off processes, the wafer-scale monolayer $MoS_2$ are patterned into FET array. For the fabrication of PSE gate dielectric, a bottle of ionic gel was prepared first using PVDF, ionic liquid EMIMTFSI and dimethyl formamide (DMF) in a 1:0.4:10 mass ratio. Then the polymer electrolyte was applied by spin coating the ionic gel on top of the entire substrate at a spinning speed of 4000 rpm for 1 min. After spin-coating, the devices were heated at 160 °C for 30 min to remove the solvent DMF and form the PSE. The whole PSE fabrication process was carried out in an argon-filled glove box. Electrical measurements were carried out with Keithley 2636B, Keithley 2400 SMU instruments and Keithley 2002 Multimeter in a four-probe vacuum station with a base pressure of ∼$10^{-5}$ mbar at room temperature. The annealing treatment is carried out in the same probe station in vacuum, heating at 100 °C for 5 min and cooling down naturally to room temperature.

*Radiation Exposure*: The radiation experiments were carried out using a $^{60}Co$ γ-ray source. All samples were mounted facing toward the γ-ray source. By controlling the distance of the sample from the $^{60}Co$ γ-ray source, the devices were irradiated with dose rates of 60.7 and 523 rad $s^{-1}$. The accumulated irradiation dose (TID) was varied by adjusting the irradiation time.

**Acknowledgements**

This project has been supported by the National Basic Research Program of China (Grant Nos. 2019YFA0308402, 2018YFA0305604), the National Natural Science Foundation of China (NSFC Grant Nos. 11934001, 11774010, 11921005, 11834017), Beijing Municipal Natural Science Foundation (Grant No. JQ20002), and Key-Area Research and Development Program of Guangdong Province (Grant No. 2020B0101340001).

**Supporting Information**

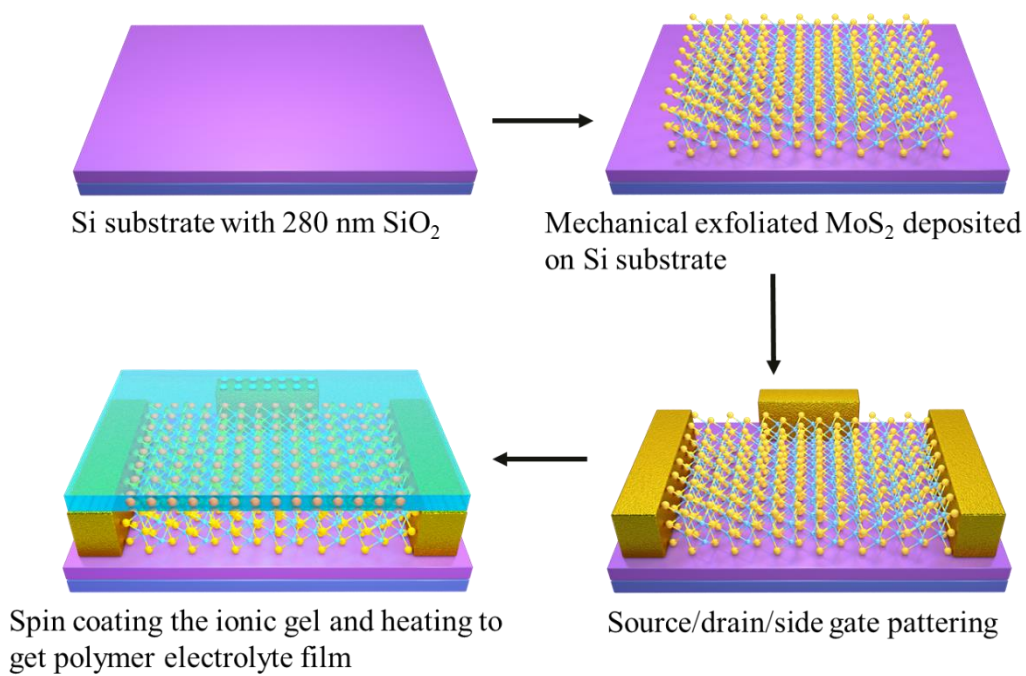

**Figure S1.** Schematic illustration of the fabrication procedure of MoS$_2$ PSE-FETs.

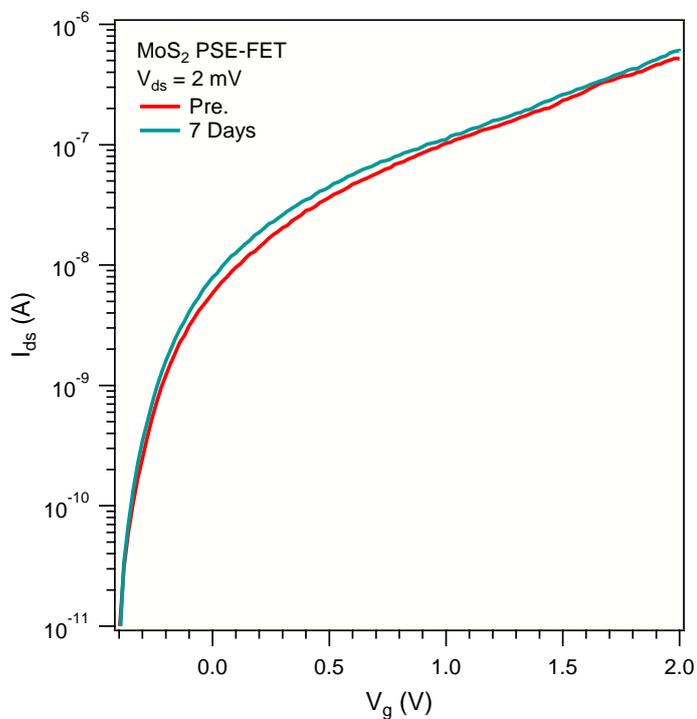

**Figure S2.** Comparison of transfer characteristics for the MoS$_2$ PSE-FET at the freshly prepared state and after being stored in air for 7 days.

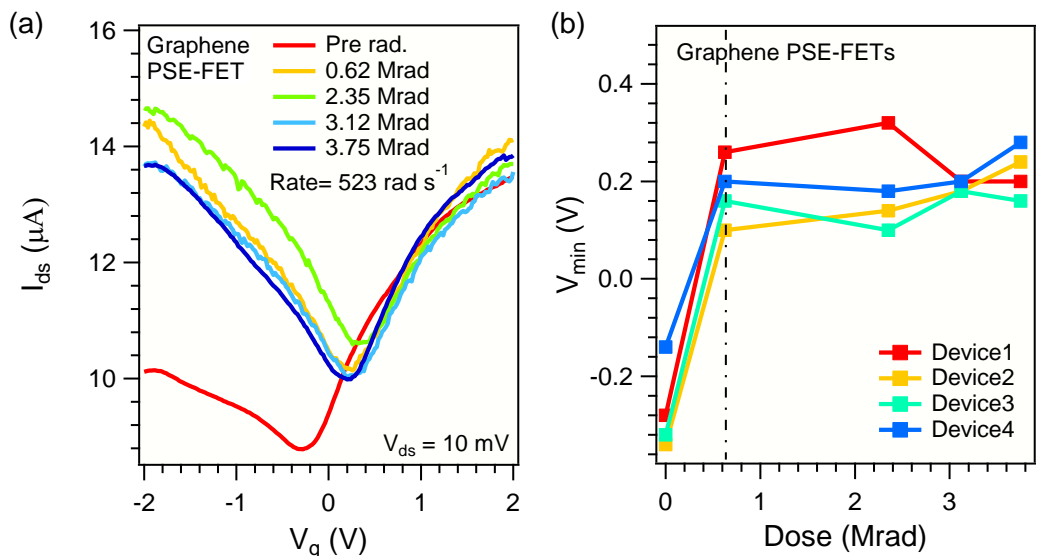

**Figure S3.** TID-dependent properties of graphene PSE-FETs. a) Transfer characteristics of the graphene PSE-FET at $V_{ds} = 10$ mV with different TIDs. b) $V_{min}$ (the $V_g$ for the minimum $I_{ds}$) of 4 graphene PSE-FETs as a function of TID.

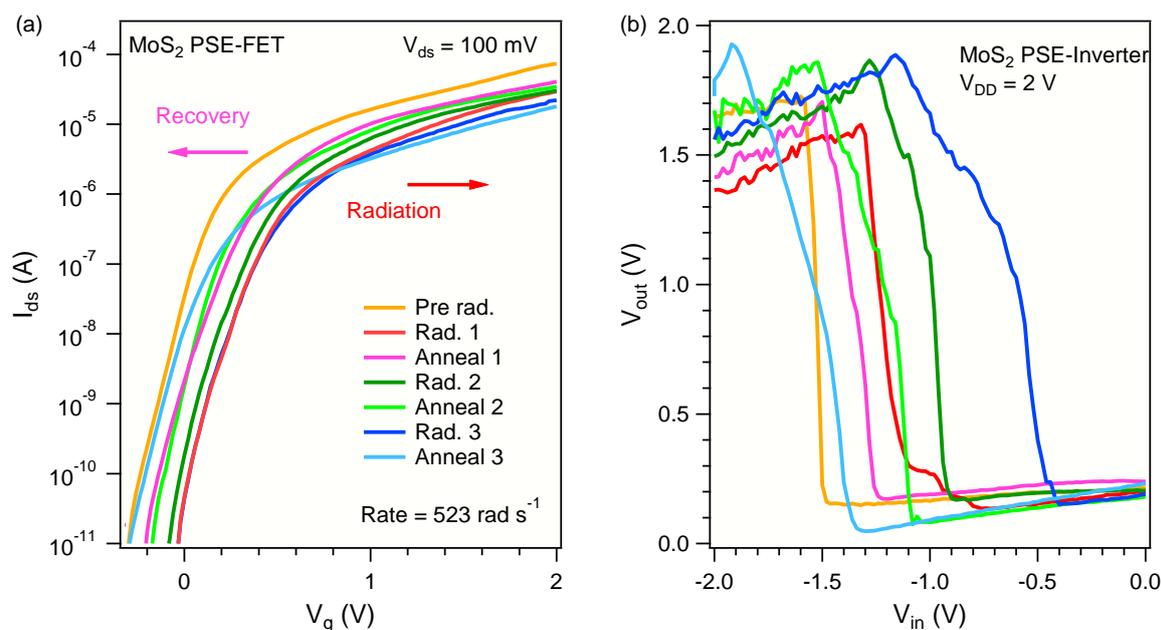

**Figure S4.** Annealing repair effects of a MoS$_2$ PSE-FET and inverter. a) Transfer characteristics of the MoS$_2$ PSE-FET with three radiation-recovery cycles. b) VTC properties of the MoS$_2$ PSE-inverter with three radiation-recovery cycles.

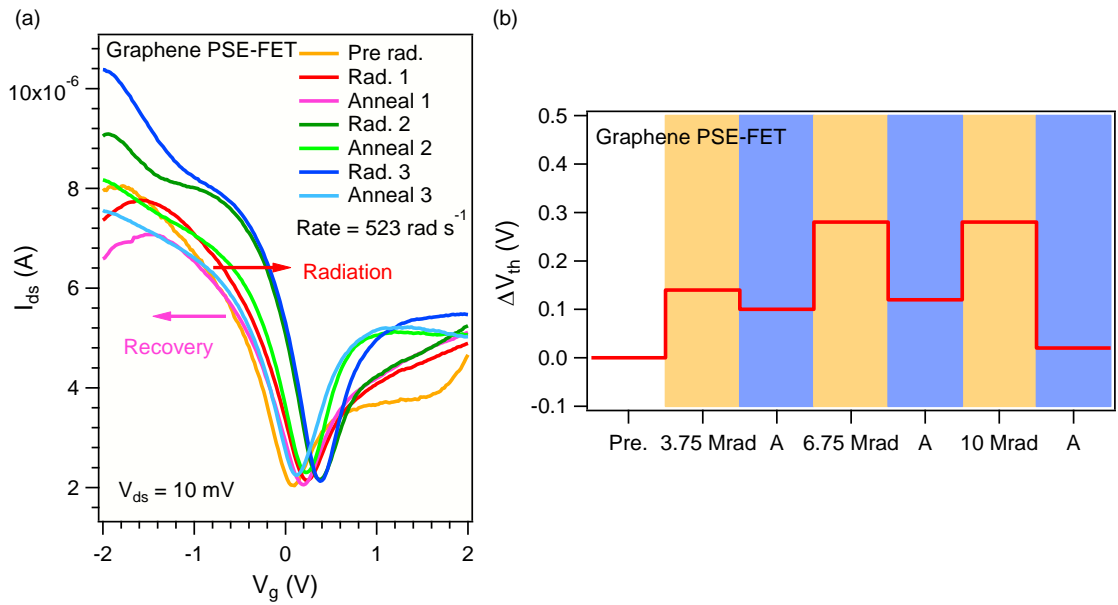

**Figure S5.** Annealing repair effects of a graphene PSE-FET. a) Transfer characteristics of the graphene PSE-FET with three radiation-recovery cycles. b) Extracted $V_{th}$ shifts for the three cycles of irradiation and annealing repair of the graphene PSE-FET.

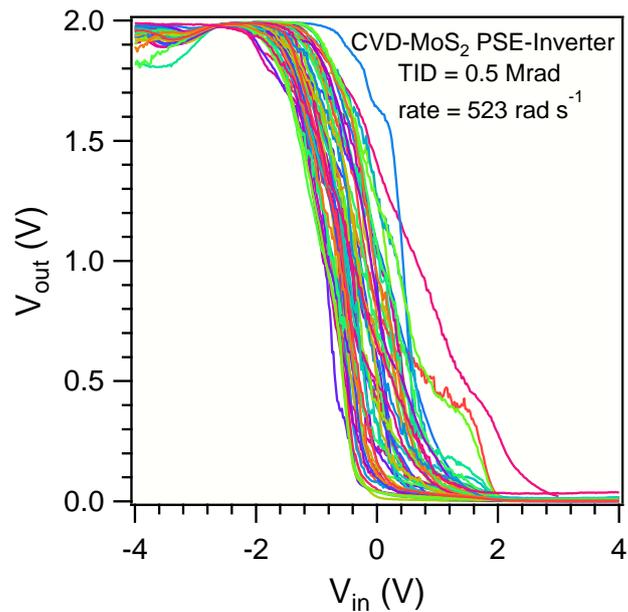

**Figure S6.** VTC curves of the 5×10 array of CVD grown MoS$_2$ PSE-inverters after 0.5 Mrad (rate = 523 rad s$^{-1}$) radiation. The histogram of the $\Delta V_{tr}$ distribution is shown in the maintext Figure 4d.

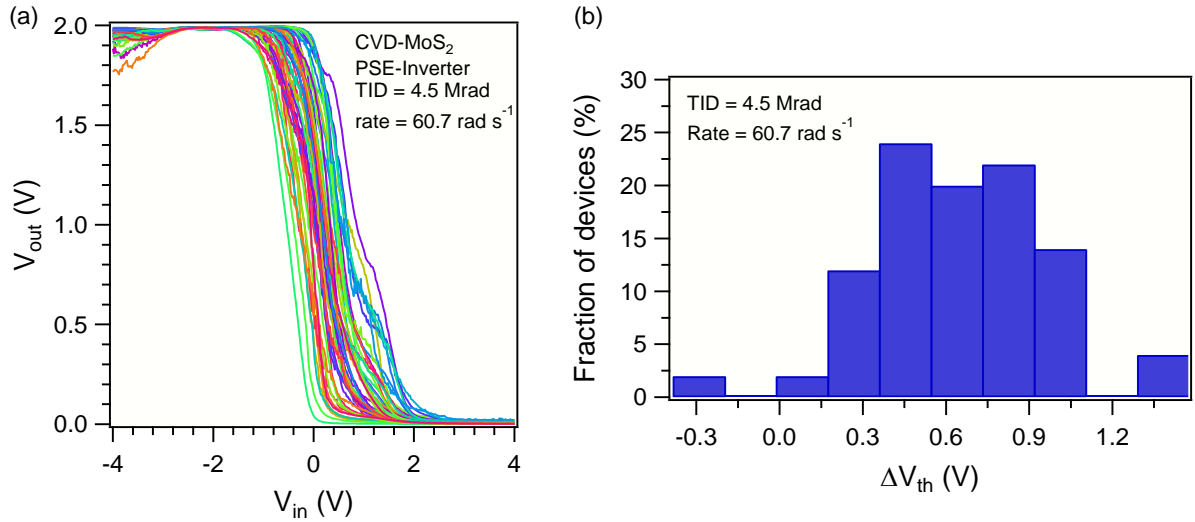

**Figure S7.** Radiation effect of wafer-scale CVD grown $MoS_2$ PSE-inverters array with a low radiation rate. a) VTC curves and b) Histogram of statistical $\Delta V_{tr}$ data of the 5×10 array of $MoS_2$ PSE-inverters after 4.5 Mrad (rate = 60.7 rad s$^{-1}$) radiation.

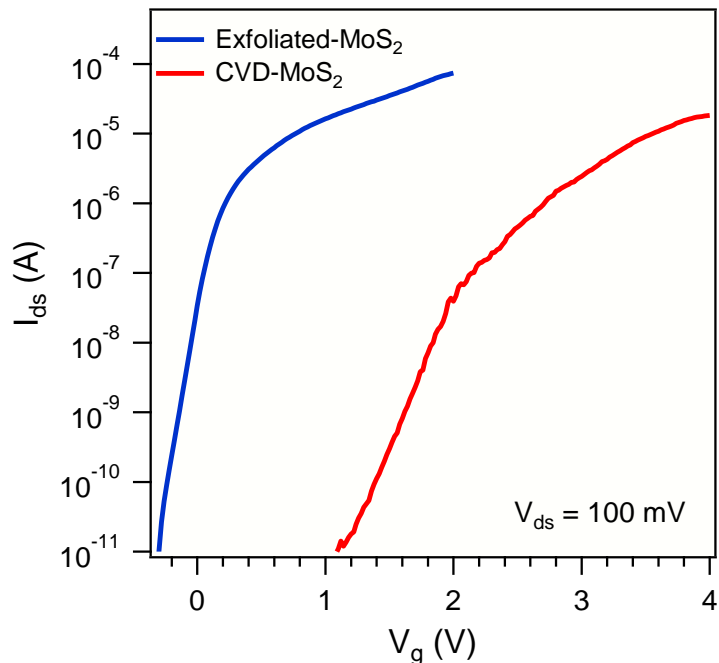

**Figure S8.** Comparison of the transfer characteristics of the PSE-FETs made of mechanically exfoliated (blue) and CVD grown (red) $MoS_2$, respectively.